\documentclass{ws-ijmpe}

\usepackage{color}

\newcommand{\px}{$\pi\!-\!\Xi$ }
\newcommand{\pp}{$\pi\!-\!\pi$ }
\newcommand{\s}{\mbox{$\sqrt{s_{NN}}$=200}~GeV} 
\newcommand{\q}{\mbox{$\sqrt{s_{NN}}$=62}~GeV}

\newcommand{\Ao}{\mbox{$A_{00}$}}
\newcommand{\Al}{\mbox{$A_{11}$}}

\begin{document}

\markboth{Petr Chaloupka}{\px correlations in d+Au and Au+Au collisions at STAR}

\catchline{}{}{}{}{}

\title{\px CORRELATIONS IN d+Au AND Au+Au COLLISIONS AT STAR}

\author{\footnotesize PETR CHALOUPKA}

\address{Nuclear Physics Institute ASCR,\\
Na Truhlarce 39/64, 18086 Praha 8, Czech Republic,\\
petrchal@rcf.rhic.bnl.gov}

\maketitle


\begin{abstract}
Qualitative comparison of source sizes from \px correlations analyses in d+Au and
Au+Au collisions at \s\ and \q\ is presented.
For the most central Au+Au collisions at \s\
we report first quantitative results concerning size of the \px source and relative shift of the
average emission points between $\pi$ and $\Xi$  showing
that the homogeneity region of $\Xi$ source is  smaller 
then that of pion and significantly shifted in the transverse direction.
\end{abstract}

\section{Non-identical particle correlations}
Important task in relativistic heavy ions research program 
carried out at RHIC
is to study the space-time evolution of the strongly interacting
matter created in the collisions. 
Measurements of momentum correlations of particles 
at small relative velocities are used to extract 
information about space-time extension of the particle-emitting source 
at the time of kinetic freeze-out\cite{Lisa_Pratt_review}.
STAR has performed high statistics femtoscopic analyses
for a wide variety of different identical and non-identical systems\cite{Debasish,Chaloupka_QM06}.
General conclusion from these measurements is that
the space-time structure of the source is strongly affected by collective
expansion of the hot and dense matter.
Non-identical particle measurements are sensitive not only to 
the size of the system, but also to an emission asymmetry 
between different particle species\cite{Lednicky} which can arise 
as a consequence of such an expansion\cite{blastwave}.
Since this effect is predicted to increase with a mass difference between
the particles, studying correlations in system, such as \px, 
where the mass difference is large, should provide rather
sensitive test of transverse expansion of the matter.

\section{Data selection and analysis technique}
STAR detector is capable of tracking charged particles at full azimuthal coverage
in  pseudorapidity window of $|\eta|<1.8$\,.
The particles are identified via
specific energy loss ({\it dE/dx}) in the gas of the main  Time Projection Chamber.
The detector is thus well suited to topologically reconstruct individual
charged $\Xi(\bar{\Xi})$-hyperons 
using decay chain \mbox{$\Xi\rightarrow\Lambda+\pi$}, and subsequently 
\mbox{$\Lambda\rightarrow\pi+p$}.
The high statistics of data collected by the STAR experiment has allowed
to carry out \px femtoscopic correlation analyses for three different data sets at two different
collision energies: d+Au and Au+Au collisions at \s, 
and Au+Au collisions at \q. 

Standard event mixing technique\cite{mercedesHBT} was used to obtain the \px
correlation function $C(\vec{k^*})$, where $\vec{k^*}=\vec{p}_{\pi}=-\vec{p}_{\Xi}$
denotes three-momentum of the first particle in the rest frame of the pair.
To remove correlations of non-femtoscopic origin the mixed pairs were 
constructed from events with sufficient proximity
in primary vertex position along the beam direction, multiplicity
and event plane orientation variables. 
Pair cuts were used to remove effects of track splitting and merging.
The raw correlation function was then corrected for 
purity of \px pairs.
Pair purity 
was calculated as a product of purities of both particle species.
While $\Xi$-purity was obtained from reconstructed $\Xi$
invariant mass plot as a function of transverse momentum,
the purity of pion sample was estimated from $\sqrt{\lambda}$
of the standard parametrization of the identical \pp correlation function\cite{Lisa_Pratt_review,mercedesHBT}.
In order to take an advantage of already analyzed STAR $\pi\!-\!\pi$ HBT data,
we applied the same cuts to the pion sample as in \cite{mercedesHBT}
and used $\lambda$ parameter from this analysis.
The same rapidity selection cut $y<|0.5|$ was also used for $\Xi$.
The particle identification method together with acceptance of the STAR detector 
allows to reconstruct $\Xi$s at mid-rapidity in the $p_{t}$ range of \mbox{[0.7, 3.]~GeV/c}.

\begin{figure}[th]
\centerline{ \psfig{file=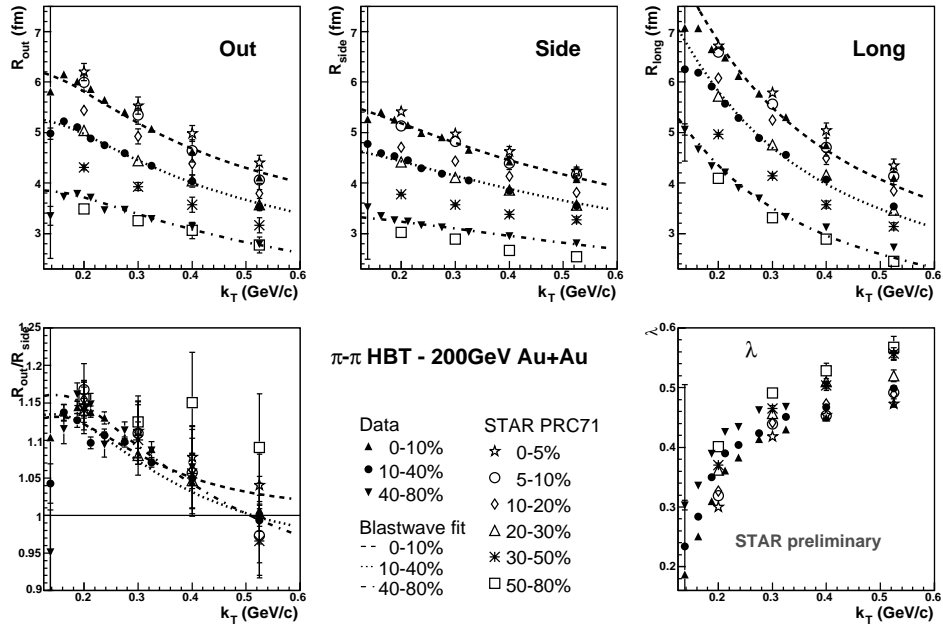,width=\textwidth}} 
\vspace*{8pt}
\caption{
Parameters of 3D Gaussian fit to the
correlation functions of identical charged pions produced in Au+Au
collisions at 200 GeV. Previous\protect\cite{mercedesHBT} (open symbols) and this analysis
(full circles) with solid lines showing Blastwave fits. 
Error bars contain only statistical uncertainties.}
\label{fig-pipiHBT}
\end{figure}

In order to further increase the number of pairs in the 200~GeV Au+Au data set
cut \mbox{$|y|<0.8$} instead of \mbox{$|y|< 0.5$} was employed for both particle species.
To use the above described pion purity calculation procedure
with new cuts, the $\pi\!-\!\pi$ HBT analysis was repeated with
high statistics data collected by the STAR experiment in RHIC Run IV
with the same event centrality and pion cuts as those used
in the \px analysis.
The results, presented in  Fig.~\ref{fig-pipiHBT} 
are consistent with previously published STAR \pp HBT analysis\cite{mercedesHBT} 
extending it to the region of low-$k_t$ which was not accessible before. 
This allows to perform more accurate  purity correction in 200~GeV Au+Au data set
done individually for each $\vec{k^*} = (k^*,\cos\theta,\varphi)$ bin of the \mbox{3-dimensional}
correlation function, as explained in section 4.

\section {1D results - source size information}
\begin{figure}[th]
\begin{center}
$\begin{array}{c@{\hspace{0.2cm}}c}
\psfig{file=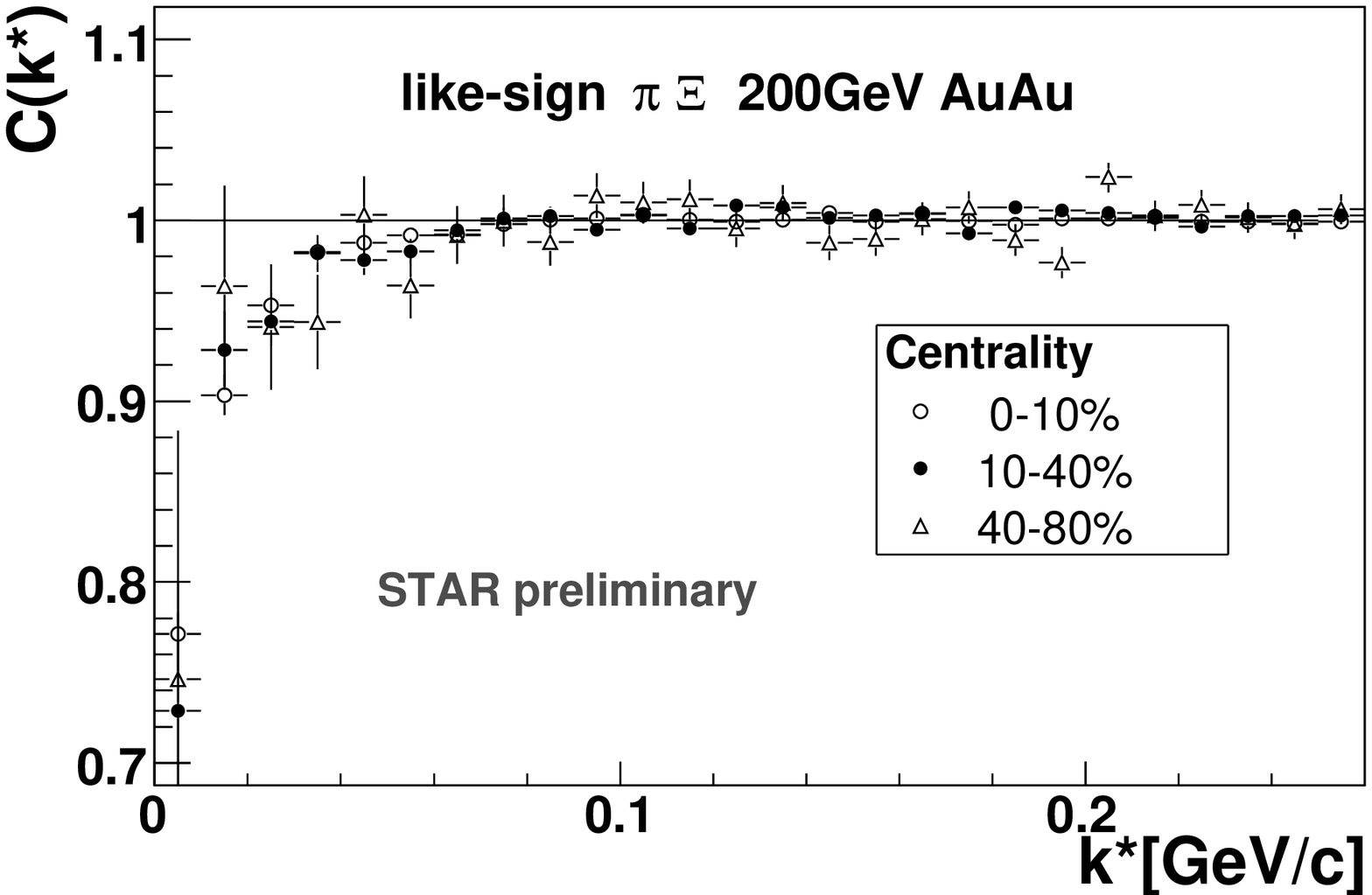,width=0.48\textwidth}& \psfig{file=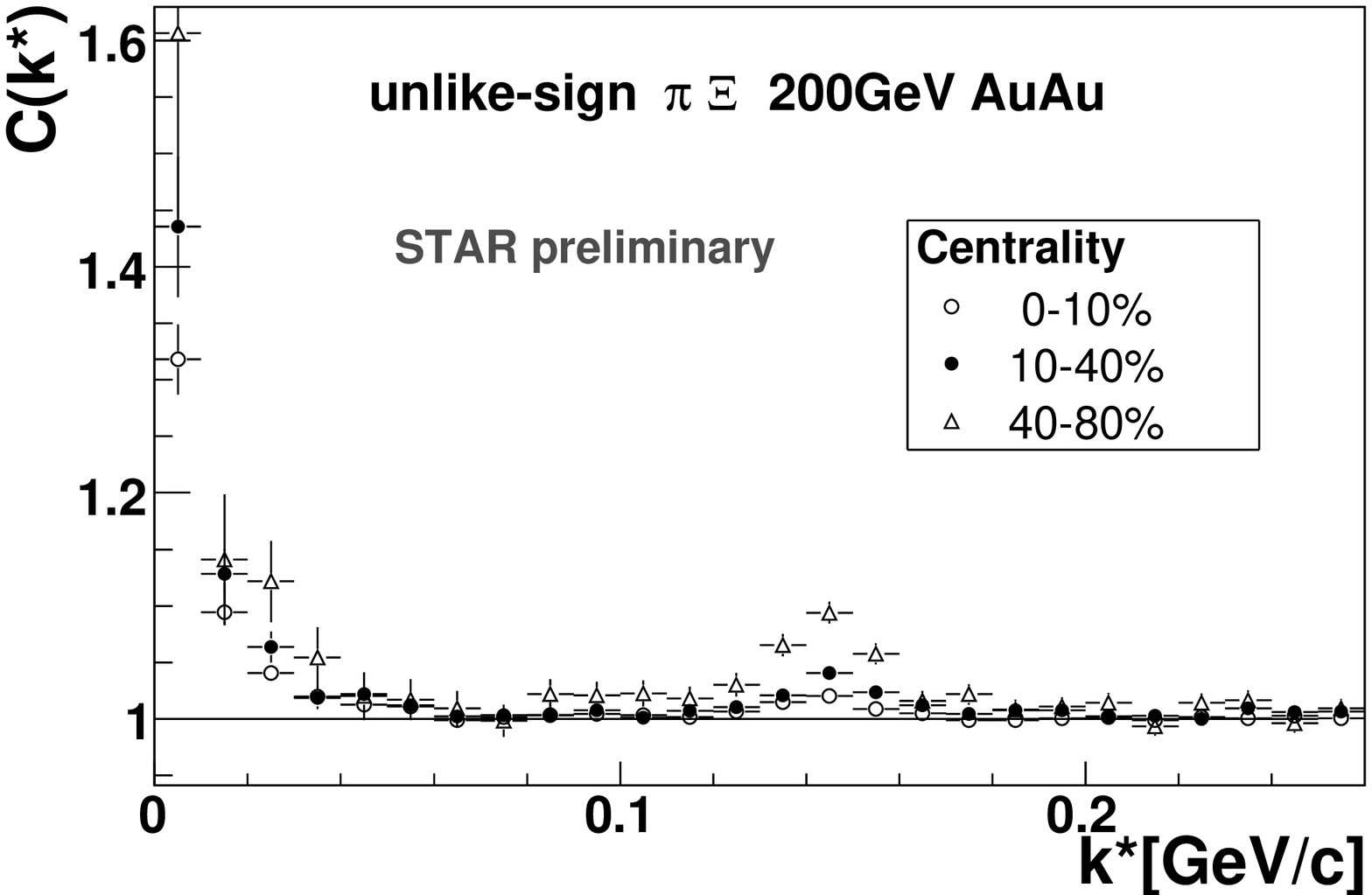,width=0.48\textwidth} \\
\mbox{\bf (a)}                           &  \mbox{\bf (b)}\\
\vspace{4pt}\\
\psfig{file=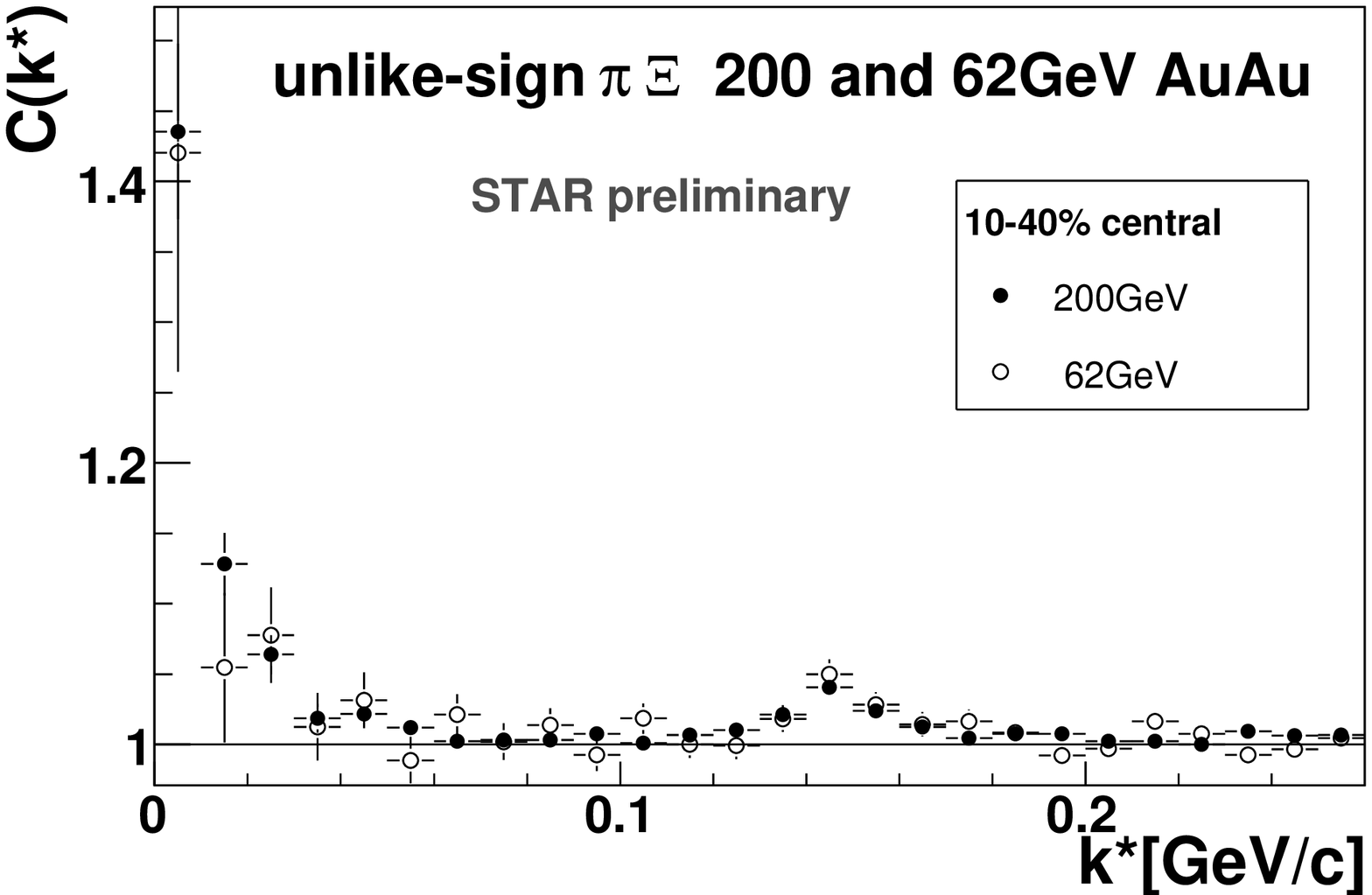,width=0.48\textwidth}   & \psfig{file=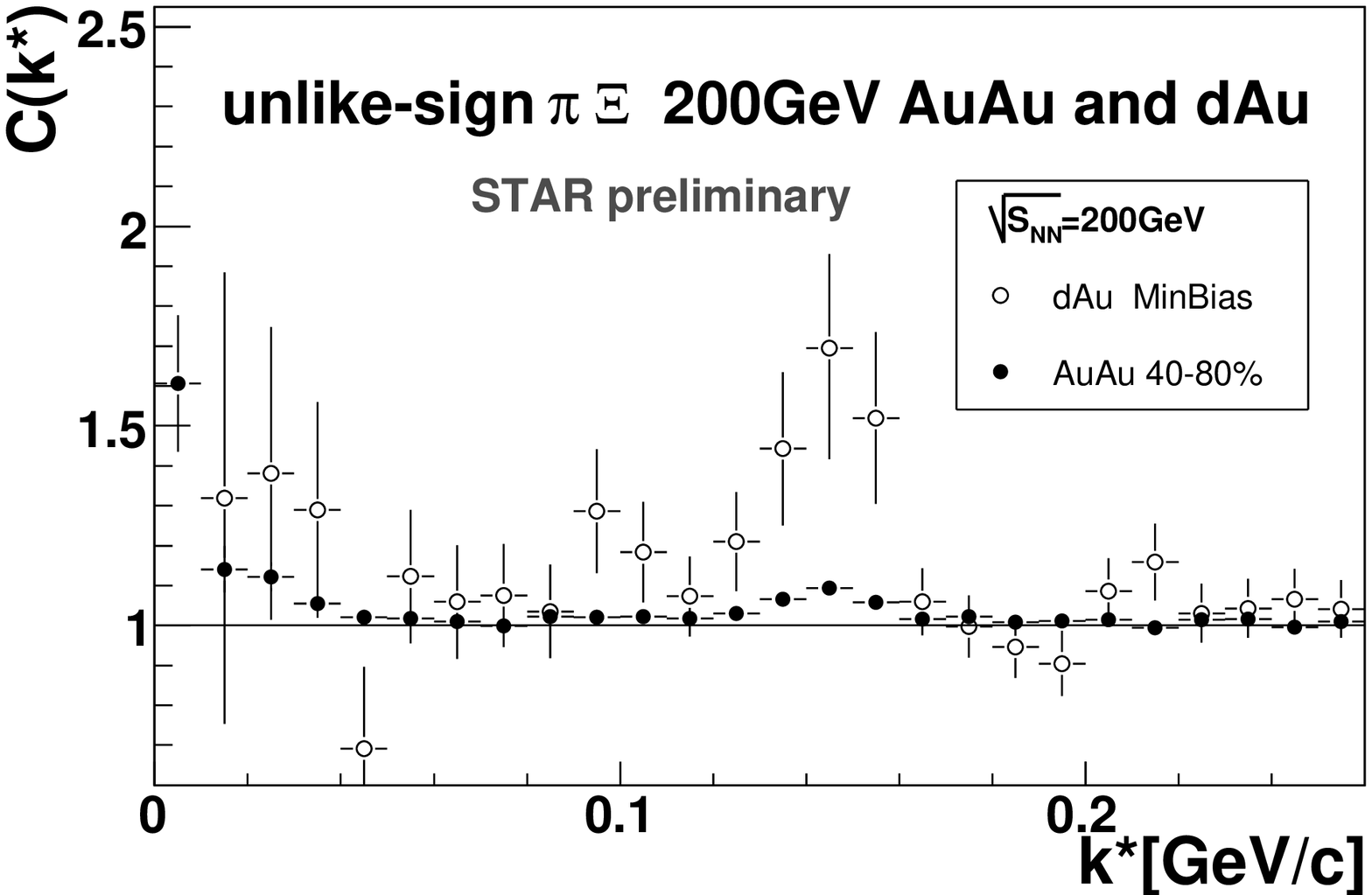,width=0.48\textwidth} \\
\mbox{\bf (c)}                           & \mbox{\bf (d)}\\
\end{array}$
\end{center}
\vspace*{8pt}
\caption{Top: The centrality dependence of the correlation function
in Au+Au collisions at \s\ of (a)like-sign and (b) unlike-sign \px pairs.
\newline
Bottom: (c) Comparison of unlike-sign correlation function for mid-peripheral events at \s\ and 62~GeV. 
(d) Correlation function for peripheral Au+Au and minimum bias d+Au at the same collision energy \s.
}
\label{fig-1D}
\end{figure}

Correlation function in terms of a single variable
 - $|\vec{k^*}|$ only - carries information about the size 
of the emission source. 
The magnitude of the correlation effect increases with decreasing source size.
The results for all analyzed systems are presented in Fig.~\ref{fig-1D}.
For all charge combinations the low-$k^*$ region is dominated
by Coulomb interaction. 
In unlike-sign charge combination	
peak at \mbox{$k^*\approx150~$MeV$/c$} is well visible. 
Since $k^*$ is connected
directly to invariant mass ($M_{inv}$) of the pair 
 \mbox{$k^*={[M_{inv}^2-(m_\pi-m_\Xi)^2]^{1/2}[M_{inv}^2-(m_\pi+m_\Xi)^2]^{1/2}}/{(2M_{inv})}$}
the peak corresponds to a $\Xi^{*}(1530)$ resonance. 
As can be seen (Fig.~\ref{fig-1D}(b)) the region of  the resonance is sensitive to the size
 of the source, and does not suffer from the low statistics like 
the Coulomb region. It can therefore be used to compare sizes of sources
that would otherwise be
impossible to compare via Coulomb part. 
In \mbox{Fig.~\ref{fig-1D}(c,d)} we
compare 200~GeV Au+Au results with data from
 200~GeV d+Au and 62~GeV Au+Au collisions.
We do not observe strong dependence
of the \px source on the collision energy. 
However, even though the d+Au  statistics is low, we can conclude 
that the source in d+Au collision is substantially smaller than that of 
most peripheral Au+Au data bin.

\section {3D results - source shift information}

\begin{figure}[th]
\begin{center}
$\begin{array}{c@{\hspace{0.2cm}}c}
\psfig{file=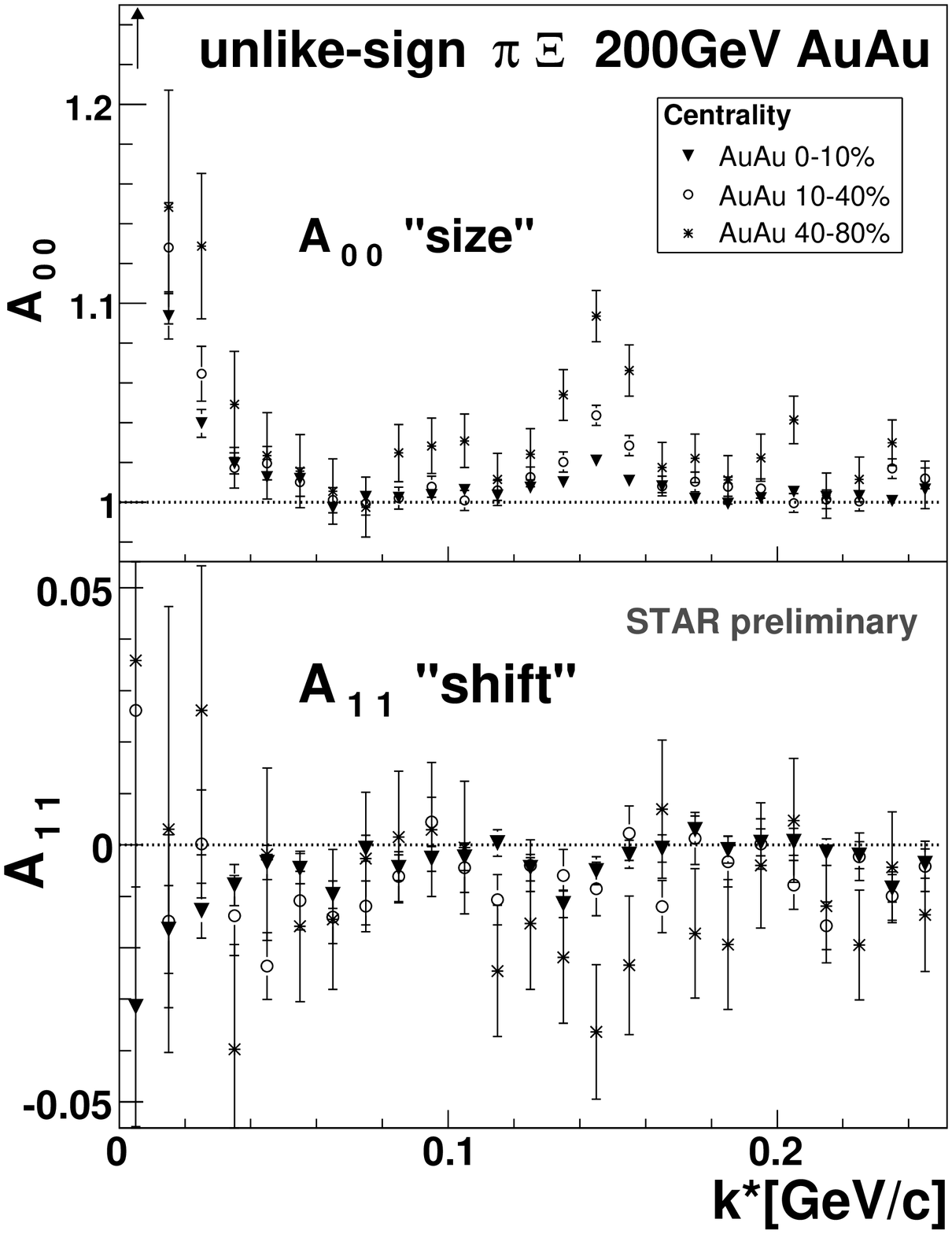,width=0.49\textwidth}& \psfig{file=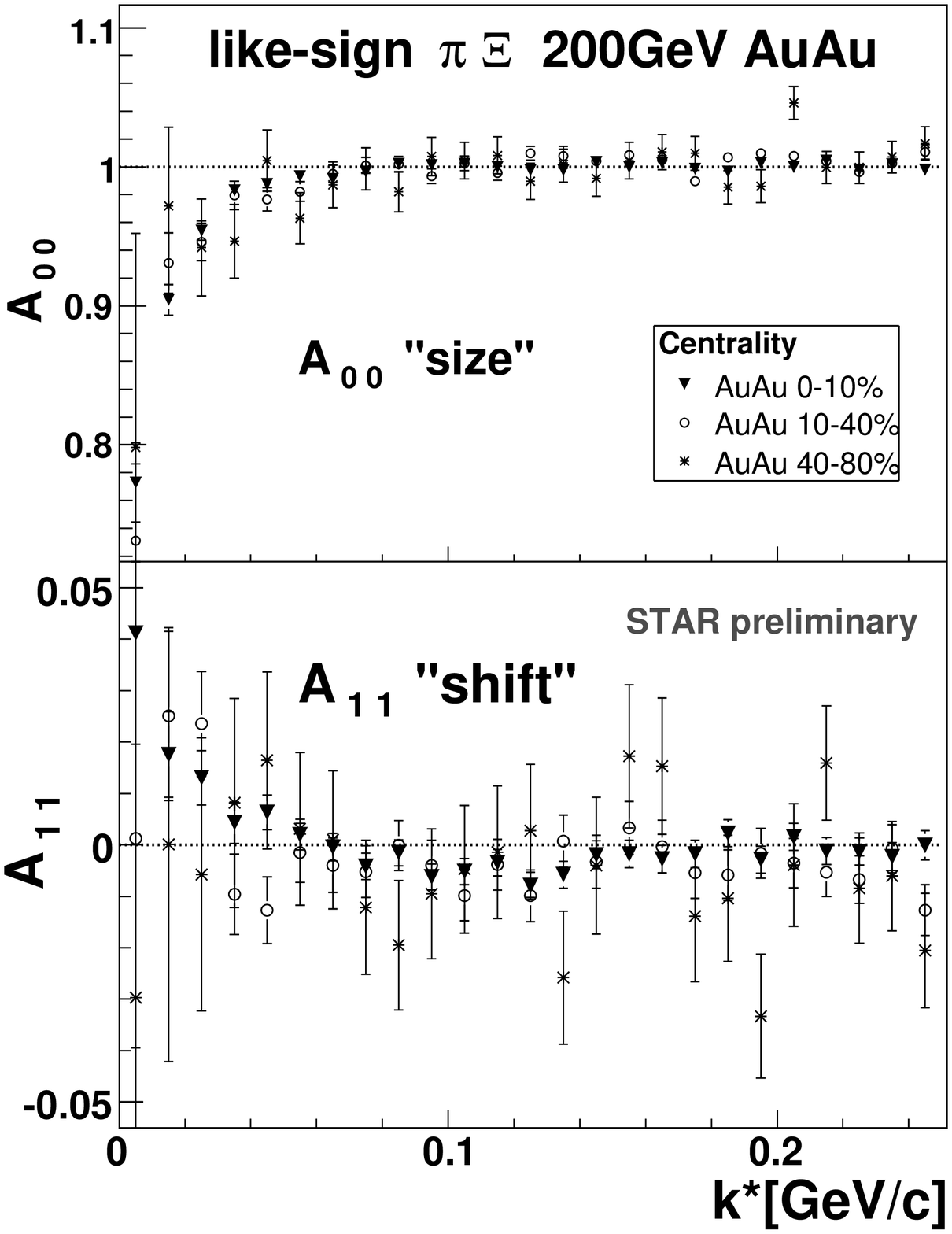,width=0.49\textwidth} \\
\end{array}$
\end{center}
\vspace*{8pt}
\caption{\Ao$(k^*)$ and \Al$(k^*)$ coefficients of spherical decomposition
of $C(\vec{k^*})$ for \mbox{(left)unlike-sign} and \mbox{(right)like-sign} \px pairs from three different
centrality bins in 200~GeV Au+Au collision.}
\label{fig-SHcent}
\end{figure}

\begin{figure}[th]
\begin{minipage}[h]{0.49\textwidth}
\centering
\psfig{file=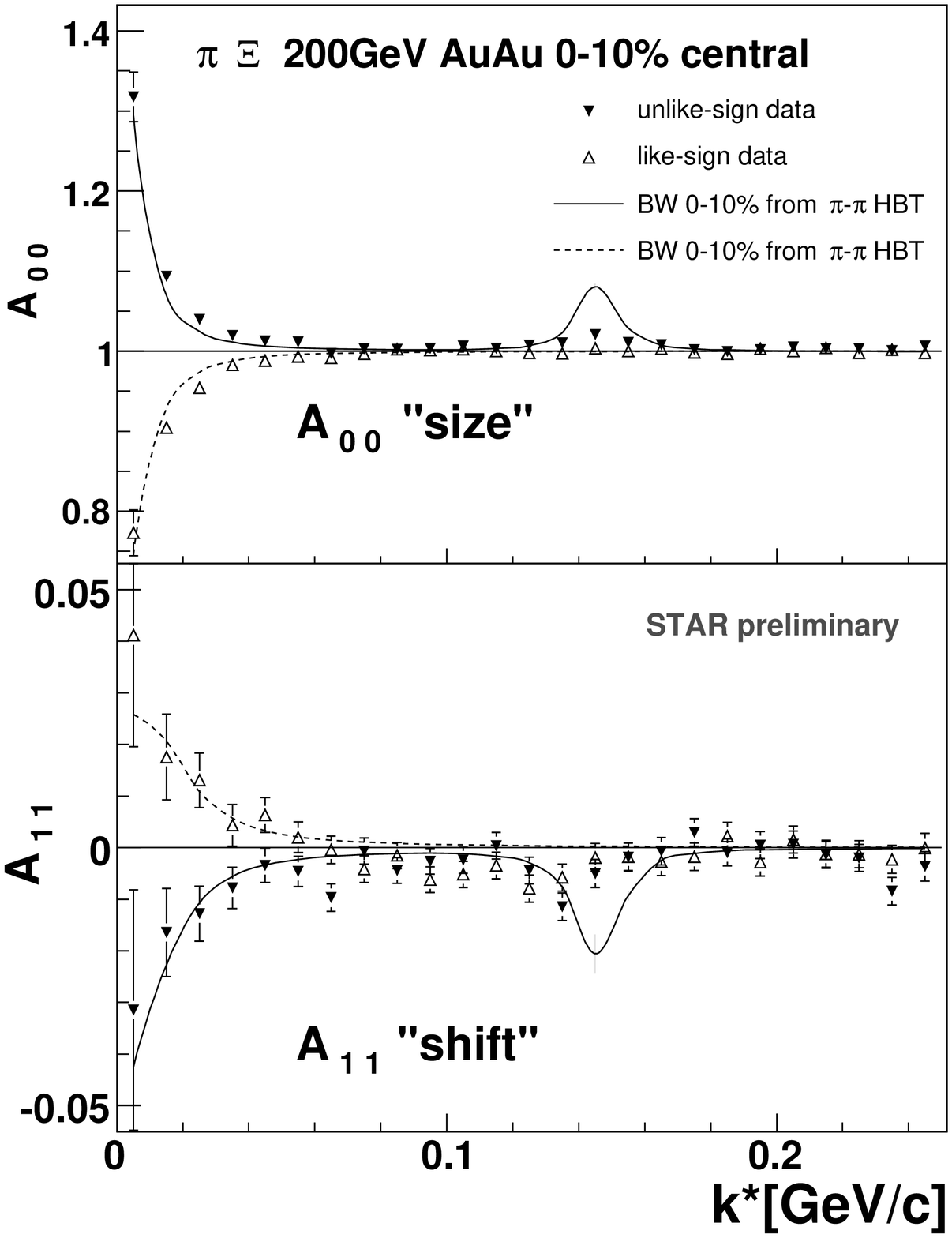,width=0.99\textwidth}
\vspace{8pt}
\caption{Comparison of \Ao$(k^*)$ and \Al$(k^*)$ coefficients 
from 10\% most central 200~GeV Au+Au collisions with the FSI model predictions.}
\label{fig-SHmodel}
\end{minipage}\hfill
\begin{minipage}[h]{0.49\textwidth}
\centering
\psfig{file=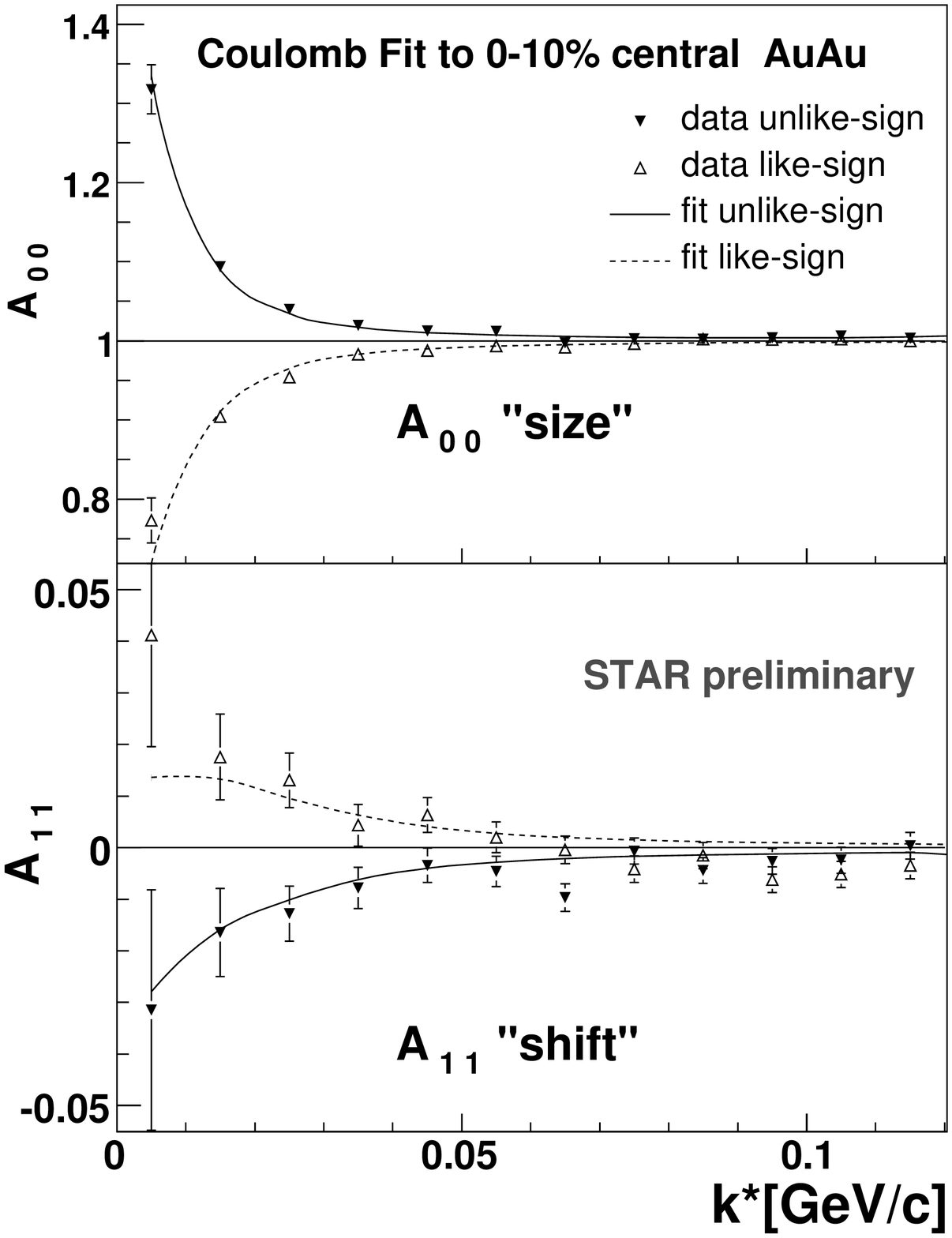,width=0.99\textwidth}
\vspace{8pt}
\caption{Fit of the low-$k^*$ part of correlation function 
in the most central 200~GeV Au+Au data using Coulomb interaction only.
\label{fig-SHfit}
\newline
}
\end{minipage}
\hfill
\end{figure}

The information about shift in the average emission points between
$\pi$ and $\Xi$ can be extracted from the angular part of the
3D correlation function \mbox{$C(\vec{k^*})=C(k^*,\cos\theta,\varphi)$}\cite{Lednicky}.
Decomposition of the correlation function into spherical
harmonics\cite{Chajecki_Lisa_SH} can be used to access such an asymmetry.
In Fig.~\ref{fig-SHcent} is shown centrality dependence
for \Ao\, and \Al\, coefficients of the spherical decomposition of $C(\vec{k^*})$
in the 200~GeV Au+Au collision.
The coefficient \Al\, which is non-zero in all centrality bins shows that
the average space-time emission point of the two particles is not the same.
The experimental results for the most central, the highest statistics,
bin are compared in Fig.~\ref{fig-SHmodel} to theoretical predictions
using calculation of Coulomb and strong final state interaction(FSI)\cite{Pratt_model}.
For this calculation 
emission coordinates of both particles were generated using Blastwave model\cite{blastwave} 
with parameters from the fit to the $\pi\!-\!\pi$ HBT results in Fig.~\ref{fig-pipiHBT}.
The same parameters were used for both $\pi$ and $\Xi$ source, thus assuming
significant flow of the $\Xi$. 
The correlation function in Coulomb region is in qualitative 
agreement with the data - the orientation of the shift and its magnitude agrees with 
the scenario in which $\Xi$ undergoes transverse expansion.
However, the  strong final state interaction calculation overpredicts both,
$A_{00}$ and $A_{11}$, coefficients.

\subsection{Coulomb fit results}
Due to the discrepancy between the Coulomb and strong interaction region
only the low-$k^*$, Coulomb dominated, part of the $C(\vec{k^*})$ was selected
for fitting,excluding the region of the $\Xi^*$ peak.
The same Gaussian parametrization as in previous STAR non-identical correlation analysis\cite{Kisiel_pion_kaon}
was assumed to describe separation distribution $\vec{r^*}$ in the pair rest frame.
Two free parameters are then the Gaussian radius
 \mbox{$\sigma=\sigma_{r^*_{out}}=\sigma_{r^*_{side}}=\sigma_{r^*_{long}}$}
and the mean
\mbox{$\langle{\Delta}r^*_{out}\rangle={\langle}r^*_{out}(\pi)-r^*_{out}(\Xi)\rangle$}
which is related to the shift in the average emission points
of the two particle species.
The theoretical correlation function was calculated using momenta
of pairs extracted from the real data. 
Emission coordinates were randomly generated using our two-parameter source description.
Values of the source parameters were extracted by finding a minimum value of $\chi^2$
between calculated and real correlation function.
Fitting was performed simultaneously for both like and unlike-sign 
correlation functions.
For most central Au+Au collision at 200GeV, see Fig.~\ref{fig-SHfit}, this method 
yields \mbox{$\sigma=(6.7\pm1.0)~fm$}, \mbox{$\langle{\Delta}r^*_{out}\rangle=(-5.6\pm1.0)~fm$}.
The errors are purely statistical. Systematic error studies are under
way and their values are expected to be of the order of the statistical ones.
Note, that the size of the Gaussian radius is approximately equal to the size of the 
pion source itself\cite{mercedesHBT}, 
showing that homogeneity region of $\Xi$ source is smaller then that of pions.
The negative value of the shift means that $\Xi$ average 
emission point is positioned more to the outside of the whole 
fire-ball than the average emission point of pions. 
Both these results are consistent with a scenario in which $\Xi$ takes part
in the rapid transverse  expansion of the system.

\section*{Acknowledgements}
This work was supported in part by the IRP AV0Z10480505
and by GACR grants 202/04/0793 and 202/07/0079.

\bibliographystyle{unsrt}
\bibliography{bib/my2.bib}

\end{document}